\documentclass[twoside,11pt]{article}

\usepackage{jmlr2e}
\usepackage[boxruled]{algorithm2e}
\SetAlFnt{\scriptsize}
\SetAlCapFnt{\scriptsize}
\SetAlCapNameFnt{\scriptsize}

\makeatletter
\renewcommand{\algocf@caption@boxruled}{%
  \hrule
  \hbox to \hsize{%
    \vrule\hskip-0.4pt
    \vbox{   
       \vskip\interspacetitleboxruled%
       \unhbox\algocf@capbox\hfill
       \vskip\interspacetitleboxruled
       }%
     \hskip-0.4pt\vrule%
   }\nointerlineskip%
}%
\makeatother

\usepackage{subcaption}

\hypersetup{
  colorlinks = true,
  urlcolor = blue, 
  linkcolor = blue,
  citecolor = blue,
  pdftitle = {Pycobra: A Python Toolbox for Ensemble Learning and Visualisation - \today}, 
  pdfauthor = {Benjamin Guedj and Bhargav Srinivasa Desikan}}

\usepackage{lastpage}
\jmlrheading{18}{2018}{1-pageref{LastPage}}{4/17; Revised 3/18}{4/18}{17-228}
{Benjamin Guedj and Bhargav Srinivasa Desikan}

\ShortHeadings{Pycobra}{Guedj and Srinivasa Desikan}
\firstpageno{1}

\def\bX{\mathbf{X}}
\def\1{\mathbf{1}}

\begin{document}

\title{Pycobra: A Python Toolbox for Ensemble Learning and Visualisation}
%pycobra: A Python toolkit for ensemble regression analysis and visualisation

\author{\name Benjamin Guedj \email benjamin.guedj@inria.fr \\
       \addr Modal project-team, Lille - Nord Europe research center\\
       Inria, France
       \AND
       \name Bhargav Srinivasa Desikan \email bhargav.srinivasa-desikan@inria.fr \\
       \addr Modal project-team, Lille - Nord Europe research center\\
       Inria, France
       }

\editor{Geoff Holmes}

\maketitle

\begin{abstract}%   <- trailing '%' for backward compatibility of .sty file
We introduce \texttt{pycobra}, a Python library devoted to ensemble learning (regression and classification) and visualisation. Its main assets are the implementation of several ensemble learning algorithms, a flexible and generic interface to compare and blend any existing machine learning algorithm available in Python libraries (as long as a \texttt{predict} method is given),
and visualisation tools such as Voronoi tessellations. \texttt{pycobra} is fully \texttt{scikit-learn} compatible and is released under the MIT open-source license. \texttt{pycobra} can be downloaded from the Python Package Index (PyPi) and Machine Learning Open Source Software (MLOSS). The current version (along with Jupyter notebooks, extensive documentation, and continuous integration tests) is available at \href{https://github.com/bhargavvader/pycobra}{https://github.com/bhargavvader/pycobra} and official documentation website is \href{https://modal.lille.inria.fr/pycobra}{https://modal.lille.inria.fr/pycobra}.
\end{abstract}


% \section{Instructions}
% \href{http://www.jmlr.org/mloss/mloss-info.html}{Instructions from JMLR}

%\tableofcontents %to be commented before submission

\vspace{-.3cm}

\section{Introduction}

Combined statistical procedures, also known as ensemble or aggregation methods, are very popular in Machine Learning competitions -- the celebrated Netflix Challenge (as are repeatedly Kaggle competitions) was won by an ensemble technique \citep[see for example][for a discussion]{bell2007lessons}. In a nutshell, ensemble methods combine different predictors (each trained on the same dataset) to improve performance. \texttt{scikit-learn} offers implementations of some ensemble methods but it does not ship with a specific analysis toolbox for aggregation. \texttt{pycobra} attempts to fill this gap by providing to the \texttt{scikit-learn} environment a toolbox to analyse and visualise the different preliminary predictors (also called machines hereafter), in addition to providing pythonic implementations of several ensemble algorithms (the COBRA algorithm introduced by \citealp{biau2016cobra}; the Exponential Weighted Aggregate (EWA) introduced by \citealp{vovk}; a COBRA-flavored majority vote inspired by \citealp{mojirsheibani}). All these algorithms are supported by oracle bounds which prove that the risk of the aggregated predictor is upper-bounded by the smallest risk of the initial pool of predictors, up to a remainder term which decays to zero (typically at a rate $\mathcal{O}(1/\sqrt{n})$ or faster).
\smallskip

COBRA (standing for COmBined Regression Alternative) is a nonlinear ensemble method designed for regression problems. 
To predict the response for a new data point, COBRA operates in two steps. First, it retains data points for which the prediction made by preliminary machines is close (in the Euclidean sense) to the prediction made for the new point. This is called the \emph{consensus step}. Second, the final prediction is then formed by averaging responses over the retained points' indices.
COBRA outputs a predictor which outperforms any combination of the preliminary predictors \cite[as shown by see Theorem 2.1 in][]{biau2016cobra}. We describe the COBRA algorithm as implemented in \texttt{pycobra} in \autoref{pycobra}.
Exponential weights have been used for a long time in statistical learning theory \citep{vovk}. The EWA algorithm implemented in Pycobra is inspired by the description of \cite{dalalyan}. EWA amounts to forming an exponentially weighted average of preliminary predictors, where the weights include a measure of each predictor's performance.
COBRA has been inspired by the work of \cite{mojirsheibani} in classification, and therefore \texttt{pycobra} includes a classification version of COBRA called \texttt{ClassifierCobra}. The consensus step shares a similar philosophy to what is done in COBRA regression but the final prediction is delivered by a majority vote among the labels of the retained data points.
\smallskip

\texttt{pycobra} allows the user to gauge the performance of the preliminary predictors used in the aggregation, with built-in methods to easily plot boxplots and QQ-plots. A salient feature of \texttt{pycobra} is using Voronoi tessellations for generic visualisation. By implementing ensemble algorithms which were not represented in the Python machine learning community, and providing a variety of tools to visualise and analyse their behavior and performance, we present to the machine learning open source community a toolbox designed for ensemble learning and visualisation.

\vspace{-.2cm}
\section{The \texttt{pycobra} library}

Our toolbox is written in Python and uses \texttt{NumPy} \citep{walt2011numpy} and \texttt{scikit-learn} \citep{pedregosa2011scikit} for computation and linear algebra operations. \texttt{Matplotlib} \citep{hunter2007matplotlib} is used for visualisation purposes, and \texttt{SciPy} \citep{jones2014scipy} is used to help create Voronoi tessellations. Tutorials and examples are created using Jupyter IPython notebooks \citep{PER-GRA:2007}. Currently \texttt{pycobra} supports any machine learning algorithm with a \texttt{predict} method, casting our procedure into a very flexible and generic framework. By default, \texttt{pycobra} relies on \texttt{scikit-learn} implementations of Lasso, Random Forest, Decision Trees and Ridge regression for the \texttt{EWA} and \texttt{COBRA} implementations. As for \texttt{ClassifierCobra}, \texttt{scikit-learn} implementations of SVM classifier, KNN, Decision Tree classifier, and a classifier which implements regularised linear models with SGD (SGDClassifier object). \texttt{pycobra} itself and all software it relies on is open source, with no dependence on proprietary software. \autoref{pycobra} presents the pseudo-code of the COBRA implementation. All the pycobra estimators are \texttt{scikit-learn} compatible and can be used as part of the existing \texttt{scikit-learn} ecosystem, such as \href{http://scikit-learn.org/stable/modules/grid_search.html}{[\texttt{GridSearchCV}]} and \href{http://scikit-learn.org/stable/modules/generated/sklearn.pipeline.Pipeline.html}{[\texttt{Pipeline}]}.
%\smallskip
While hyperparameter initialisation is systematically done using \texttt{scikit-learn}'s \texttt{GridSearchCV}, \texttt{pycobra}'s \texttt{Diagnostics} class allows us to compare between different combinations of the constituent predictors and data-splitting, among other basic parameters.

\begin{algorithm}[t]
\caption{The original COBRA algorithm from \cite{biau2016cobra}.}
 \KwData{input vector $\mathbf{X}$, \texttt{epsilon}, \texttt{alpha}, \texttt{basic-machines}, \texttt{training-set-responses}, \texttt{training-set}}
 \# \texttt{training-set} is the set composed of all \texttt{data\_point} and the responses. \texttt{training-set-responses} is the set composed of the responses.
 
 \KwResult{prediction $\mathbf{Y}$}
 \For{\texttt{machine} $j$ in \texttt{basic-machines}}{
 \texttt{machine\_set} = { } \; \# \emph{\texttt{machine\_set} is a dictionary mapping each machine $M$ to a set \texttt{machine\_set}$[M]$ }\;
 \texttt{pred} = $r_j (\bX)$; $\qquad$ \# \emph{where $r_j(x)$ denotes the prediction made by machine $j$ at point $x$}\;
 \For {\texttt{response} in \texttt{training-set-responses}}{
   if $|\texttt{data\_point} - \texttt{pred}| \le \texttt{epsilon}$, collect \texttt{response} in \texttt{machine\_set}$[M]$ \;
   }
 }
 array = [ ] \; 
 \For {\texttt{data\_point} in \texttt{training-set}}{ 
   if at least \texttt{alpha} machines out of $M$ have collected \texttt{data\_point} in their \texttt{set}, store point in array \;
  }
 result = average(array) \;
\label{pycobra}
\end{algorithm}
%\vspace{-.2cm}

The \texttt{visualisation} class allows the user to compare all the machines used to create aggregates, as well as visualise the results, for all \texttt{pycobra} estimators. \texttt{pycobra} ships with \href{https://github.com/bhargavvader/pycobra/blob/master/notebooks/visualise.ipynb}{[a notebook on visualisation]} to illustrate this. 
\vspace{-.2cm}

\begin{figure}[b]
\begin{subfigure}{0.5\linewidth}
    \centering
    \includegraphics[width=7.5cm, height=4.8cm]{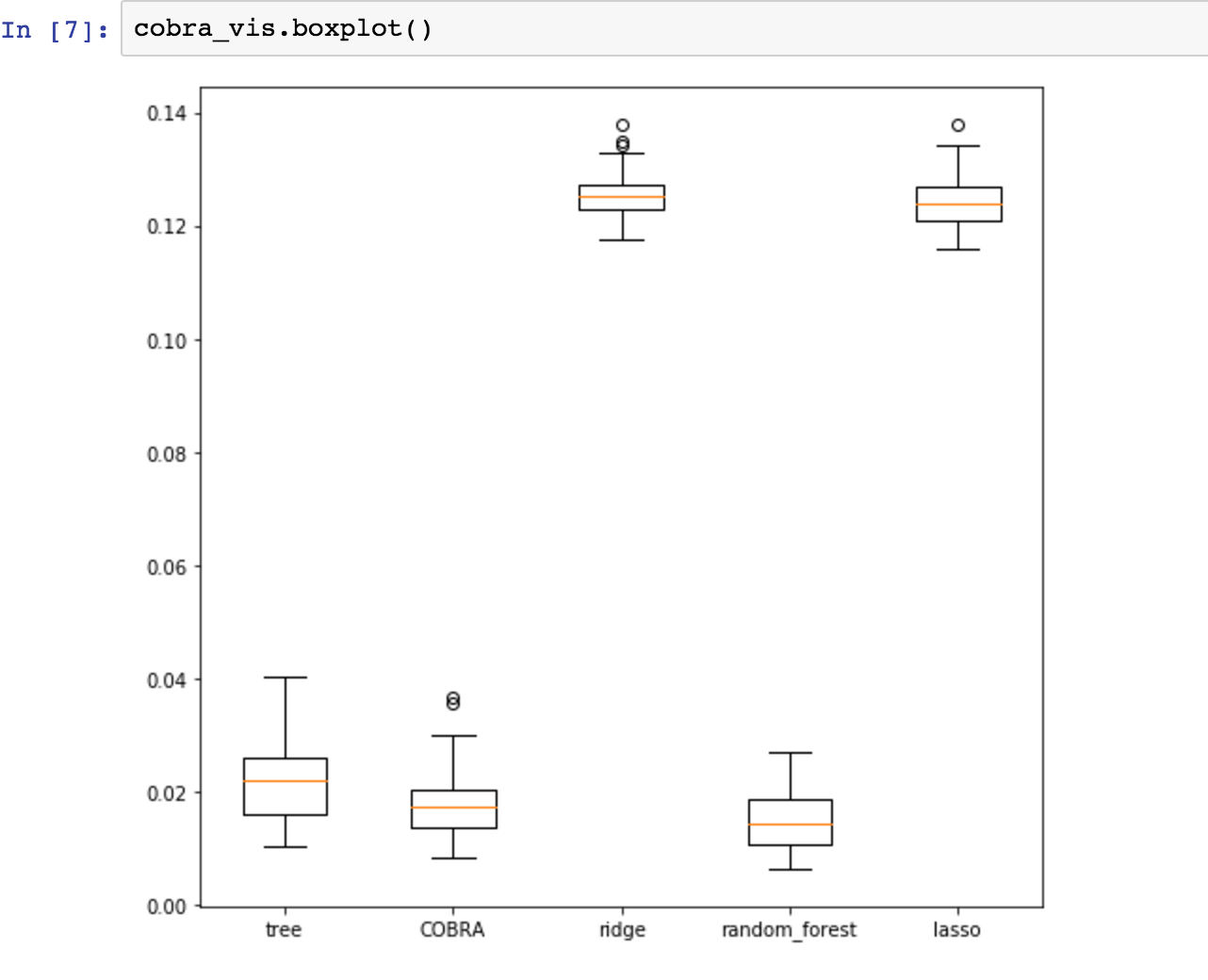}
    %\label{}
 \end{subfigure}
\begin{subfigure}{0.5\linewidth}
    \centering
    \includegraphics[width=7.5cm, height=4.8cm]{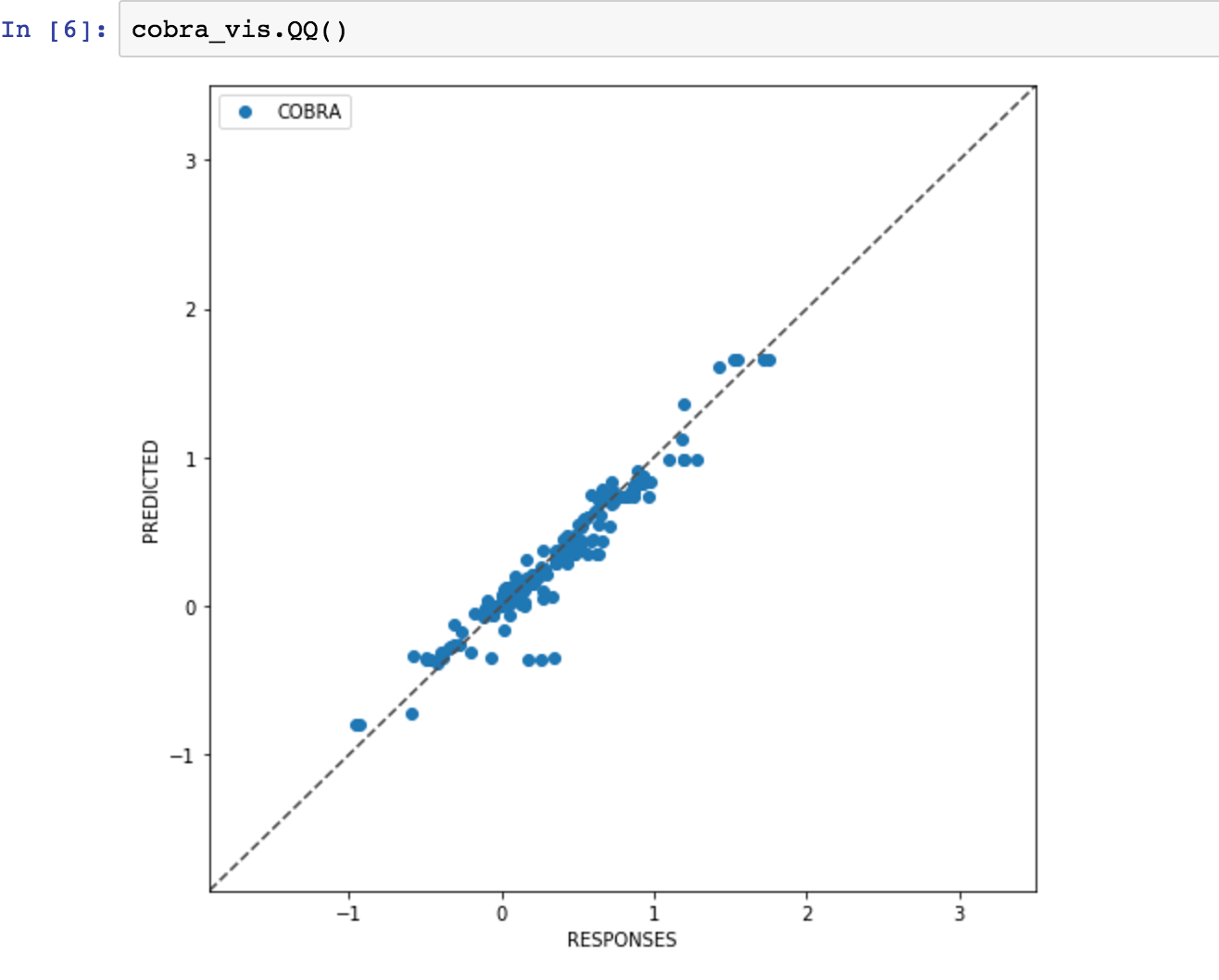}
    %\label{}
 \end{subfigure}
 \caption{Assessing the performance of regression machines and COBRA.}
 \label{fig1}
 \end{figure}

\paragraph{QQ-plots and boxplots.} Once all the basic machines are initialized and trained, the user can easily compare their performance with boxplots and QQ-plots. All the plotting details are handled by both the \texttt{diagnostics} and \texttt{visualisation} classes. An example is given in \autoref{fig1}.
\vspace{-.25cm}

\begin{figure}[h]
  \begin{subfigure}{0.5\linewidth}
    \centering
    \includegraphics[width=7cm, height=4.5cm]{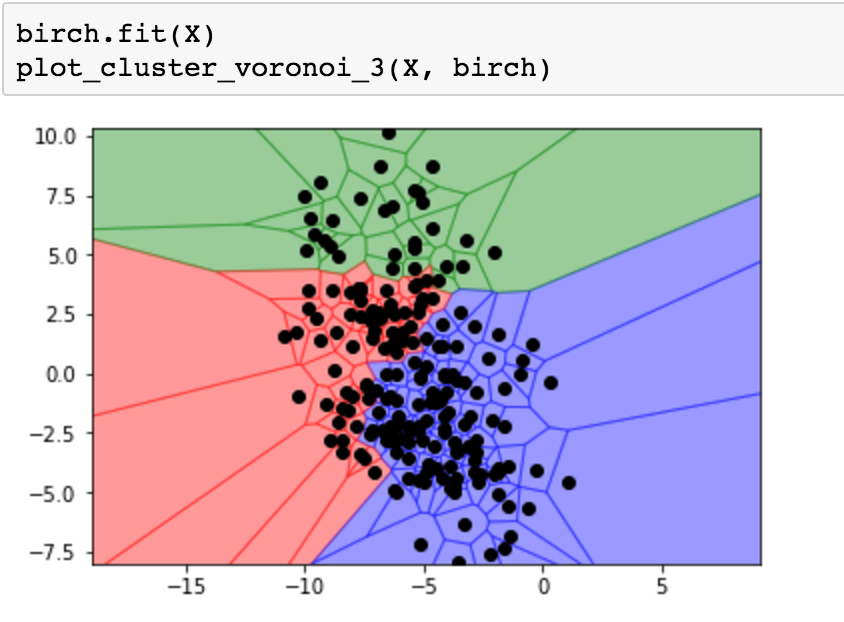}
    %\caption{}
    %\label{}
  \end{subfigure}%\hfill
  \begin{subfigure}{0.5\linewidth}
    \centering
    \includegraphics[width=7cm, height=4.5cm]{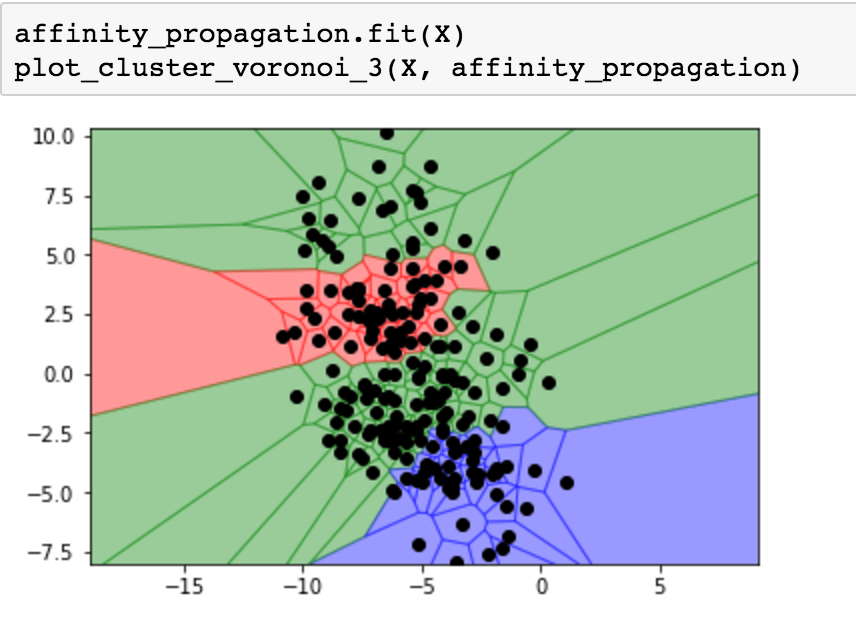}
    %\caption{}
    %\label{}
  \end{subfigure}
  \begin{subfigure}{0.5\linewidth}
    \centering
    \includegraphics[width=7cm, height=4.5cm]{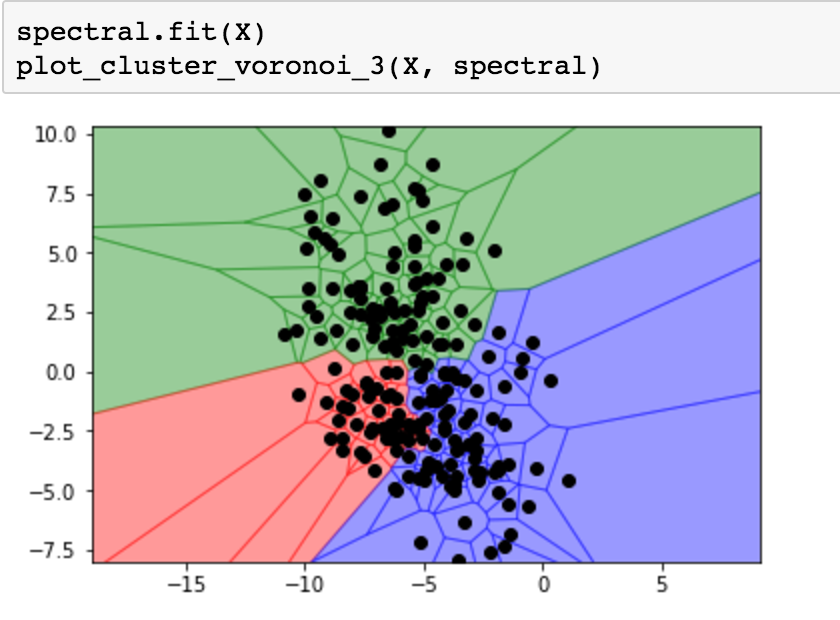}
    %\caption{}
    %\label{}
  \end{subfigure}%\hfill
  \begin{subfigure}{0.5\linewidth}
    \centering
    \includegraphics[width=7cm, height=4.5cm]{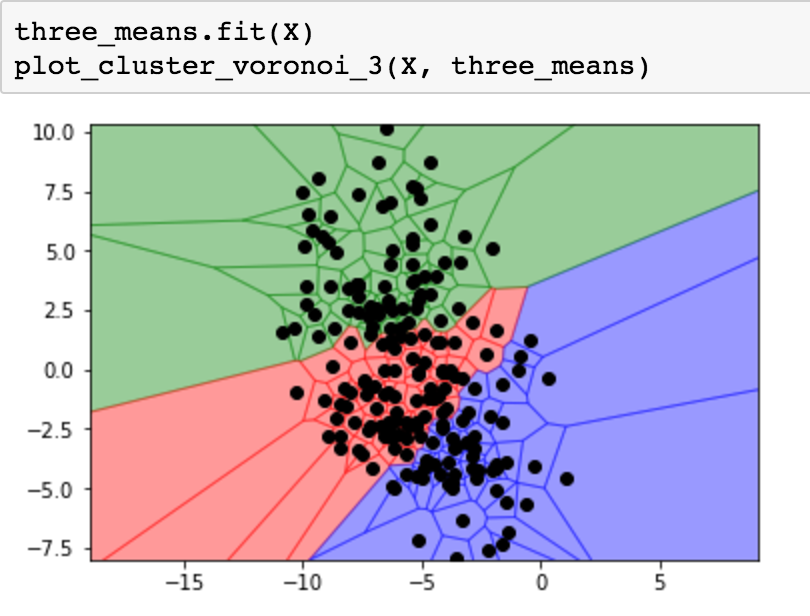}
    %\caption{}
    %\label{}
  \end{subfigure}
    \caption{Visualising clustering through Voronoi tessellations.}
      \label{fig2}
\end{figure}

\paragraph{Visualisation through Voronoi tessellations.} 
Voronoi tessellations can be used to visualise the selection of optimal machines for COBRA, as well as visualising outputs from clustering algorithms, as illustrated by the \href{https://github.com/bhargavvader/pycobra/blob/master/notebooks/voronoi_clustering.ipynb}{[tutorial notebook]}. An example is reproduced \autoref{fig2}.

\vspace{-.4cm}
\section{Project Focus}

\paragraph{Community-based development.} We intend \texttt{pycobra} to be an ongoing collaborative project. To that matter, it is referenced on \href{https://pypi.python.org/pypi/pycobra/}{[Python Package Index (PyPi)]} and \href{http://mloss.org/software/view/672/}{[Machine Learning Open Source Software]}.
Moreover, \texttt{pycobra} is under active development and available on \href{https://github.com/bhargavvader/pycobra}{[GitHub]} to promote collaborative programming, issue tracking and idea discussions.
\vspace{-.6cm}

\paragraph{Documentation and Jupyter Notebooks.} A consistent API documentation is provided, along with an additional installation guide and examples. The documentation website is available at \href{https://modal.lille.inria.fr/pycobra}{https://modal.lille.inria.fr/pycobra}. The \href{https://github.com/bhargavvader/pycobra/tree/master/notebooks}{[notebooks directory]} contains Jupyter  notebooks which serve as both a documentation tool and a tutorial resource. These notebooks cover use-cases of Pycobra, from solving regression and classification problems to using Voronoi tesselations.
\vspace{-.2cm}

\paragraph{Ease of use and quality assurance.} Ensemble learning with \texttt{pycobra} is as simple as loading trained \texttt{scikit-learn} machines -- or any machine which has a \texttt{predict} method. Visualising involves little to no parameters, and after loading machines it is straightforward to analyse their performance on a particular dataset. In order to ensure code quality, a set of unit tests is provided for all classes in \texttt{pycobra}, and continuous integration via \href{https://travis-ci.org}{[Travis CI]} ensures all commits are tested. The package follows the PEP8 convention, keeping the code easy to read and contribute to.

\vspace{-.4cm}

\section{Conclusion and Future Work}

The future of \texttt{pycobra} would be to grow its user base by adding new ways to add predictors, as well as further implement ensemble learning techniques and visualisation tools. Statistical aggregation and ensemble learning are an important part of the machine learning literature and are widely used by practitioners, yet it seems  under-represented in the machine learning open source community. By creating \texttt{pycobra} and releasing it to the community, we intend to enrich the existing ecosystem with ensemble algorithms, a generic toolbox for ensemble learning on-the-go, along with analysis and visualisation tools.

\bibliography{biblio}

\end{document}